\documentclass[floats,floatfix,showpacs,amssymb,prd,twocolumn,superscriptaddress,nofootinbib]{revtex4-1}

\usepackage{graphicx,epsf, epsfig, amssymb}
\usepackage{bm}
\usepackage{longtable}
\usepackage{color}
\usepackage[breaklinks]{hyperref}
\usepackage{amsfonts,amsmath,amssymb,mathrsfs}

\def\be{\begin{equation}}
\def\ee{\end{equation}}
\def\beq{\begin{eqnarray}}
\def\eeq{\end{eqnarray}}

\begin{document}

\title{Effects of Post-Newtonian Spin Alignment on the Distribution of Black-Hole Recoils}

\author{Emanuele Berti}
\email{berti@phy.olemiss.edu}
\affiliation{Department of Physics and Astronomy, The University of
Mississippi, University, MS 38677, USA}
\affiliation{California Institute of Technology, Pasadena, CA 91109, USA}

\author{Michael Kesden}
\email{mhk10@nyu.edu}
\affiliation{Center for Cosmology and Particle Physics, New York University, 4 Washington Pl., New York, NY 10003, USA}

\author{Ulrich Sperhake}
\email{sperhake@tapir.caltech.edu}
\affiliation{Institut de Ci{\'e}ncies de l'Espai (CSIC-IEEC), Facultat die Ci{\'e}ncies, Campus UAB, E-08193 Bellaterra, Spain}
\affiliation{Department of Physics and Astronomy, The University of
Mississippi, University, MS 38677, USA}
\affiliation{California Institute of Technology, Pasadena, CA 91109, USA}
\affiliation{Centro Multidisciplinar de Astrof{\'i}sica -- CENTRA, Departamento de F{\'i}sica, Instituto Superior T{\'e}cnico -- IST, 1049-001 Lisboa, Portugal}


\date{\today}

\begin{abstract}
  Recent numerical relativity simulations have shown that the final
  black hole produced in a binary merger can recoil with a velocity as
  large as $5,000~{\rm km/s}$.  Because of enhanced gravitational-wave
  emission in the so-called ``hang-up'' configurations, this maximum
  recoil occurs when the black-hole spins are partially aligned with
  the orbital angular momentum. We revisit our previous statistical
  analysis of post-Newtonian evolutions of black-hole binaries in the
  light of these new findings. We demonstrate that despite these new
  configurations with enhanced recoil velocities, spin alignment
  during the post-Newtonian stage of the inspiral will still
  significantly suppress (or enhance) kick magnitudes when the initial
  spin of the more massive black hole is more (or less) closely
  aligned with the orbital angular momentum than that of the smaller
  hole. We present a preliminary study of how this post-Newtonian spin
  alignment affects the ejection probabilities of supermassive black
  holes from their host galaxies with astrophysically motivated mass
  ratio and initial spin distributions.  We find that spin alignment
  suppresses (enhances) ejection probabilities by $\sim 40\%$ (20\%)
  for an observationally motivated mass-dependent galactic escape
  velocity, and by an even greater amount for a constant escape
  velocity of 1,000~km/s. Kick suppression is thus at least a factor
  two more efficient than enhancement.
\end{abstract}

\maketitle 

\section{Introduction} \label{S:intro}

Supermassive black holes (SBHs) reside at the centers of most large
galaxies. The masses of these SBHs are tightly correlated with the
luminosity \cite{Kormendy1995}, mass \cite{Magorrian1998etal}, and
velocity dispersions \cite{Gebhardt2000} of the spheroidal components
of their host galaxies.  Large galaxies form hierarchically from the
merger of smaller galaxies, and SBH mergers are expected to accompany
the mergers of their hosts.  The final stages of these SBH mergers are
driven by the emission of copious amounts of gravitational radiation.
Conservation of linear momentum implies that the final black holes
produced in SBH mergers must recoil with linear momentum equal in
magnitude and opposite in direction to that of the anisotropically
emitted gravitational radiation.  Early estimates of these
gravitational recoils or ``kicks'' using post-Newtonian (PN)
techniques \cite{Fitchett1983} and black-hole perturbation theory
\cite{Favata2004} suggested that they would not exceed several hundred
km/s in magnitude.  More recently, progress in numerical relativity
(NR) \cite{Pretorius2005a,Campanelli2006,Baker2006} has allowed
relativists to simulate the mergers of highly spinning,
comparable-mass black holes.  For non-spinning binaries, ensuing
studies identified a maximum recoil of $175~{\rm km/s}$
\cite{Gonzalez2007}.  Simulations of spinning black holes, however,
resulted in one of the greatest surprises of numerical relativity; for
equal-mass binaries with opposite spins in the orbital plane
gravitational recoils can approach 4,000 km/s
\cite{Gonzalez2007a,Campanelli2007}, greater than the escape velocity
of even the most massive galaxies.  This theoretical result seems at
first difficult to reconcile with observations indicating that almost
all large galaxies host SBHs.

One solution to this problem is to align the SBH spins before merger
into configurations that lead to small recoils.  Gravitational
radiation extracts energy and angular momentum from the orbit of the
binary SBHs, causing them to merge on a timescale $t_{\rm GR} \propto
r^4$, where $r$ is the binary separation.  This timescale becomes
longer than the age of the universe at separations $r \sim 1$~pc,
implying that some mechanism other than gravitational radiation is
required to escort the SBHs to this separation (the ``final-parsec
problem'') \cite{1980Natur.287..307B}.  One such mechanism is the
transfer of orbital angular momentum from the SBH binary to
surrounding gas.  This gas forms a circumbinary disk about the SBHs,
and can be transferred through the gap onto accretion disks
about the individual SBHs \cite{1996ApJ...467L..77A}.  If the angular
momentum of the circumbinary disk is misaligned with the SBH spins,
the Lense-Thirring effect causes inclined annuli in the individual
accretion disks to differentially precess about the SBH spins.
Bardeen and Petterson \cite{1975ApJ...195L..65B} showed that viscous
dissipation causes this differentially precessing gas to settle into
the equatorial planes of the SBHs.  On longer timescales, these warped
accretion disks torque the SBH spins into alignment with the orbital
angular momentum of the gas at large radii, presumably that of the
circumbinary disk from which both accretion disks are being fed
\cite{2009MNRAS.399.2249P}.  Bogdanovi\'{c} et al.
\cite{Bogdanovic2007} suggested that this alignment could reduce
gravitational recoils to less than 200~km/s, in contrast to the $\sim
4,000$~km/s recoils expected for SBHs in the ``superkick''
configuration (spins in opposite directions lying in the equatorial
plane) \cite{Gonzalez2007a,Campanelli2007}.

Dotti et al. \cite{2009MNRAS.396.1640D} tested this suggestion by
performing a series of N-body smoothed particle hydrodynamics (SPH)
simulations of two $4 \times 10^6 M_\odot$ SBHs inspiraling due to
dynamical friction exerted by a $10^8 M_\odot$ circumnuclear disk in
their orbital plane.  One of the SBHs began at the center of the
circumnuclear disk, while the second SBH spiraled inwards from an
initial separation of 50~pc to a final separation of $\sim 10$~pc.
Gas particles within the Bondi-Hoyle radii \cite{1944MNRAS.104..273B}
of the SBHs were accreted, and assumed to fuel warped accretion disks
as described in Perego et al. \cite{2009MNRAS.399.2249P}.  On a
timescale of $\lesssim$~1-2~Myr, the SBH spins became aligned to
within $10^\circ$ ($30^\circ$) of their orbital angular momentum for a
cold (hot) circumnuclear disk \cite{Dotti:2009vz}.  If the partially
aligned spin configurations found in these simulations were preserved
as the SBHs inspiraled and merged, the median recoils predicted by NR
simulations would be $\lesssim 100$~km/s for dimensionless spins
$\boldsymbol{\chi}_{i}$ ($i=1,2$) with an initially uniform
distribution of magnitudes $\chi_i\in [0,1]$.

The assumption that SBH spin distributions remain unchanged between $r
\simeq 10~{\rm pc} \simeq 10^8 R_S~(M_{\rm BH}/10^6 M_\odot)^{-1}$
(where the SPH simulations end) and $r \simeq 5~R_S$ (where the NR
simulations begin), needs further examination.  Here $R_S = 2GM_{\rm
  BH}/c^2$ is the Schwarzschild radius of a SBH of mass $M_{\rm BH}$.
Because of the steep scaling $t_{\rm GR} \propto r^4$ mentioned
previously, the SBHs decouple from their circumbinary disk and spiral
inwards purely under the influence of gravitational radiation at a
binary separation $r_{\rm dec} \simeq 10^2 - 10^3 M$
\cite{1980Natur.287..307B}, where $M=m_1+m_2$ is the total mass of the
binary. Schnittman \cite{Schnittman2004} integrated PN equations of
motion and spin precession from an initial binary separation $r_i =
1000 M$ to a final separation $r_f = 10 M$ (in geometrical units where
$G = c = 1$) and discovered that partially aligned SBH spin
distributions during this portion of the inspiral are strongly
influenced by the presence of spin-orbit resonances.

These resonances are special spin configurations in which both SBH
spins and the orbital angular momentum jointly precess at the same
frequency in a common two-dimensional plane.  For binary SBHs with
mass ratio $q \equiv m_2/m_1 \leq 1$ and dimensionless spins
$\boldsymbol{\chi}_{i}$, at each separation $r$ there exist two
one-parameter families of spin-orbit resonances: one with $\Delta \phi
= 0^\circ$ and the other with $\Delta \phi = 180^\circ$, where $\Delta
\phi$ is the angle between the components of $\boldsymbol{\chi}_1$ and
$\boldsymbol{\chi}_2$ perpendicular to the orbital angular momentum
$\mathbf{L}$.  The first of these families ($\Delta \phi = 0^\circ$)
has $\theta_1 < \theta_2$, where $\theta_i\equiv
\arccos(\boldsymbol{\hat{\chi}}_i\cdot \mathbf{\hat{L}})$, while the second
family ($\Delta \phi = 180^\circ$) has $\theta_1 > \theta_2$.  As SBHs
inspiral due to the emission of gravitational radiation, they can
become captured into nearby spin-orbit resonances, substantially
altering the SBH spin distributions between $r_i$ and $r_f$.

In previous work, we examined the influence of this PN portion of the
inspiral on the expected distribution of SBH final spins
\cite{Kesden2010} and recoils \cite{Kesden2010a}.  We found that if
the spin of the more massive SBH was more closely aligned with the
orbital angular momentum ($\theta_1 < \theta_2$), the SBHs were more
likely to be captured into the $\Delta \phi = 0^\circ$ family of
resonances.  This alignment of the SBH spins prior to merger causes
them to add constructively, enhancing the spin of the final black
hole.  Symmetry requirements imply that gravitational recoils are
suppressed when the SBH spins are aligned with each other prior to
merger \cite{Boyle2007a,Boyle2007b}, so this PN alignment also reduces
the predicted distribution of recoils.  The opposite is true when
$\theta_1 > \theta_2$; SBHs are preferentially captured into the
$\Delta \phi = 180^\circ$ family of resonances, reducing the final
spins and enhancing the gravitational recoils.

Recent equal-mass NR simulations by Lousto and Zlochower
(\cite{Lousto2011,2012arXiv1201.1923L}; henceforth LZ) have
qualitatively and quantitatively changed the predicted dependence of
gravitational recoils on the SBH spins $\boldsymbol{\chi}_{i}$.  The
maximum possible kick, extrapolated to maximal initial spins, is now
$\sim 4,900$~km/s, and this kick occurs not in the previously
described ``superkick'' configuration but for spins with $\theta_i
\simeq 50^\circ$.  In light of these new findings, it is worth
investigating whether PN spin alignment still has as dramatic an
effect on the expected recoil distribution as we found previously.
This investigation is the subject of this paper.  In
Section~\ref{S:PN} we revisit our analysis of how PN evolution affects
binaries that are evolved from an initial separation of $r_i = 1000 M$
down to a final separation of $r_f = 10 M$.  We clarify the reason why
spin-orbit resonances are also effective at suppressing kicks
consistent with the new LZ results.  In Section~\ref{S:param} we set
up a more extensive set of Monte Carlo simulations of SBH
inspirals. We focus on a specific choice for the initial distribution
of SBH parameters $\{ q, \boldsymbol{\chi}_1, \boldsymbol{\chi}_2 \}$,
and justify this choice based on astrophysical considerations.  In
Section~\ref{S:res} we present the final distributions of recoil
velocities resulting from our Monte Carlo simulations, and our
predictions for the fraction of remnant SBHs that are ejected from
their host galaxies.  We summarize these results and discuss the
sources of uncertainty in our analysis in Section~\ref{S:disc}.

\section{Post-Newtonian Spin Alignment} \label{S:PN}

The recent LZ simulations \cite{Lousto2011,2012arXiv1201.1923L} have
two potentially important consequences from an astrophysical point of
view. First of all, the maximum possible ``hang-up'' kick magnitude is
about 25\% larger than predicted by the ordinary, ``no hang-up''
superkick formula: 4,900~km/s rather than 3,750~km/s. Secondly, the
largest kick occurs for a configuration in which the spins are
partially aligned with the orbital angular momentum.  Because of the
gas-dynamical alignment arguments summarized above, this configuration
seems more likely than the superkick configuration.  We would
therefore expect an enhancement in the statistical likelihood of
gravitational recoils ejecting remnant SBHs from galaxies as reported
by LZ.  We confirm this enhancement, but show that even in hang-up
kick scenarios spin-orbit resonances can significantly change the
likelihood of large recoil velocities as the SBHs inspiral from
sub-parsec scales to about $10M$.

The largest ``superkick'' recoil velocities are obtained for nearly
equal-mass SBHs.  Furthermore, the LZ kick formula is most reliable
for mass ratios close to unity, since all 48 of the new NR simulations
were carried out for $q=1$ \cite{2012arXiv1201.1923L}.  We therefore
begin our study of the likelihood of very large kicks by considering
comparable-mass binaries.  Resonance effects are not present for $q =
1$, but astrophysical binaries are not expected to be precisely equal
in mass.  For this reason we have performed Monte Carlo PN evolutions
of 900 black hole binaries with mass ratio close to but not exactly
equal to one (we choose $q=9/11$ to facilitate comparisons with
Schnittman \cite{Schnittman2004} and our own earlier work
\cite{Kesden2010,Kesden2010a}).  We considered binaries where the two
black holes have the same dimensionless spin magnitude
$\chi_1=\chi_2=\chi$, and selected five equally-spaced values of
$\chi$ in the range $[0.5, 1.0]$.  For each of the five values of
$\chi$ we further selected six equally-spaced initial values of
$\theta_1$ in the range [$5^\circ$, $30^\circ$] and six values
in the range [$150^\circ$, $175^\circ$], for a total of sixty
Monte Carlo simulations.  Each simulation is started by assuming
uniform distributions in $\Delta\phi\in [0\,,2\pi]$ and $\cos
\theta_2\in[-1\,,1]$.  We begin at an initial separation $r_i=1000M$
and use the PN equations of motion to evolve the binaries down to a
final separation $r_f=10M$ (cf.~Section~II of \cite{Kesden2010} for
details of the PN evolution).

\begin{figure}[t!]
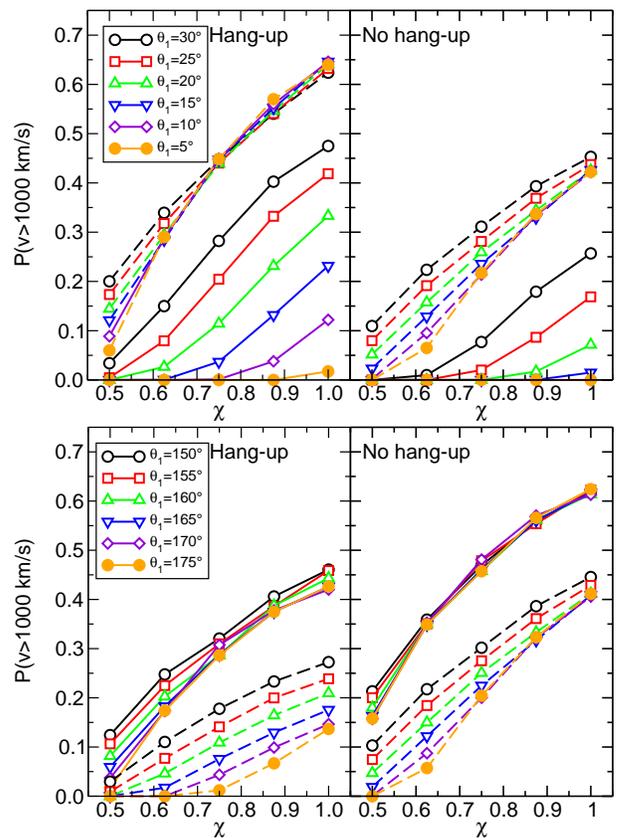

\begin{center}
\epsfig{clip=true,file=fig1a.eps,width=8.0cm}
\epsfig{clip=true,file=fig1b.eps,width=8.0cm}
\end{center}
\caption{Probability of having a recoil velocity $v>1,000$~km/s as a
  function of dimensionless spin magnitude $\chi_1 = \chi_2 = \chi$
  for binaries with mass ratio $q=9/11$.  The initial value of
  $\theta_1$ (the alignment angle of the larger SBH) is indicated by
  the different symbols, while the spin $\boldsymbol{\chi}_2$ of the
  smaller SBH is initially isotropically distributed.  Left panels:
  prediction according to the LZ papers
  \cite{Lousto2011,2012arXiv1201.1923L}; right panels: prediction
  without this newly discovered hang-up effect.  In all plots the
  dashed lines correspond to the initial distributions at $r_i =
  1000M$, while the solid lines give the distributions at $r_f = 10M$,
  after PN spin alignment has occurred.}
\label{fig:P1000}
\end{figure}

To quantify the effect of PN spin alignment, we compute the recoil
velocity distribution that would result from applying kick formulae to
our binaries {\em before} the PN evolution (at $r_i=1000M$) and close
to merger (at $r_f=10M$).  We calculate the component $v_{||}$ of the
recoil velocity parallel to $\mathbf{L}$ using both the new LZ
``hang-up'' formula \cite{Lousto2011,2012arXiv1201.1923L} and the
older kick formula from Campanelli et al. \cite{Campanelli2007}.  We
calculate the components $v_m$, $v_\bot$ perpendicular to $\mathbf{L}$
from the fitting formula of \cite{Campanelli2007}: cf.~Eq.~(1) in
\cite{Kesden2010a}.  We then compute the probability that the total
recoil velocity is greater than $v_{\rm ej}=1,000$~km/s for each of
our sixty Monte Carlo simulations (twelve values of $\theta_1$ and five
values of $\chi$).  This value of $v_{\rm ej}$ is comparable to the
escape velocity from a giant elliptical galaxy \cite{Merritt2004}; we
will consider more realistic escape velocities that depend on the host
galaxy's mass in Section~\ref{S:res}.  The results are plotted in
Figure~\ref{fig:P1000}.

\begin{figure*}[t!]
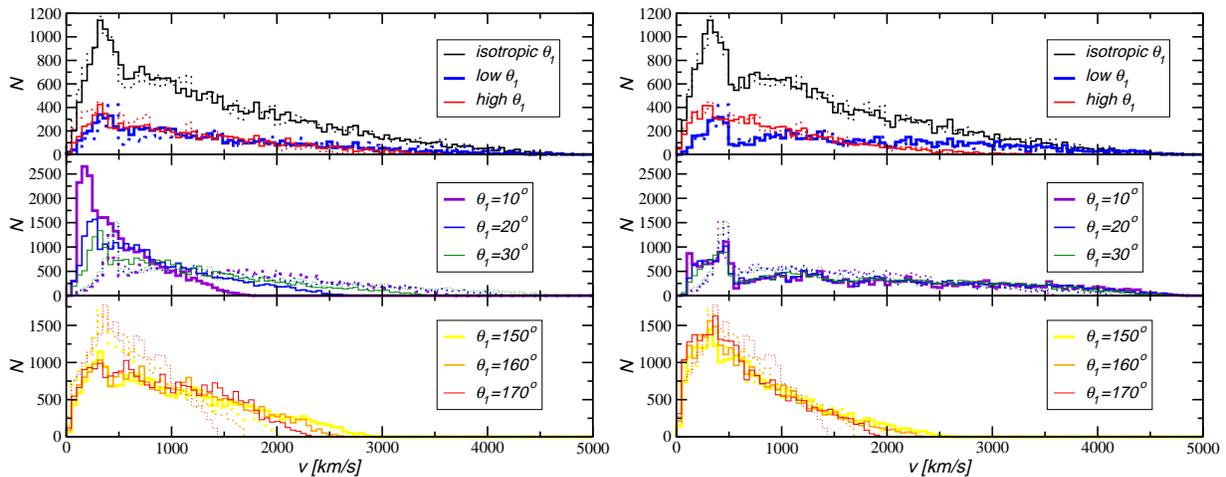

\begin{center}
\epsfig{clip=true,file=fig2a.eps,width=8.0cm}
\epsfig{clip=true,file=fig2b.eps,width=8.0cm}
\end{center}
\caption{Left panel: histograms of the recoil velocity $v$ for
  maximally spinning $(|\boldsymbol{\chi}_i|=1)$ BH mergers with mass
  ratio $q=9/11$. Dotted lines show the recoils expected if the BHs
  merge with the parameters specified at initial separation $r_i=1000
  M$. Solid curves show the expected kicks if the BHs precess as
  described in Section~2 of \cite{Kesden2010a} from $r_i$ to $r_f = 10
  M$ prior to merger. Right panel: same as in the left panel, except
  that here the solid histograms show the velocity distribution
  obtained when $\Delta \phi$ is reset to a flat distribution $\Delta
  \phi \in [0, 180^\circ]$ after the binaries reach $r_f$.}
\label{fig:deltaphi}
\end{figure*}

Figure~\ref{fig:P1000} confirms LZ's conclusion that the ``hang-up''
effect increases ejection probabilities if the spin of the heavier
companion is partially aligned with the orbital angular momentum
($\theta_1\le 30^\circ$) whereas the ejection probabilities are
reduced for partial antialignment of $\chi_1$ ($\theta_1\geq
150^\circ$); cf.~their Fig.~2.  By comparing the dashed lines in the
upper left and right panels of Figure~\ref{fig:P1000}, we see that, if
we ignore the effect of resonances (as the authors of the LZ papers
did), ejection probabilities for the aligned cases $\theta_1 \le
30^\circ$ are only mildly dependent on $\theta_1$ and roughly double
because of the hang-up kick effect.  These ejection probabilities can
be larger than $50\%$ for spins $\chi\gtrsim 0.8$ and $q=9/11$.  On
the other hand, a comparison of dashed and solid lines in each panel
shows that ejection probabilities are dramatically reduced in the
aligned case if we apply the kick formula (as we should) close to
merger, rather than at large separations. The relevance of resonant
effects depends on $\theta_1$: if the initial angle between the larger
black hole and the orbital angular momentum $\theta_1<10^\circ$,
recoil velocities larger than $v_{\rm ej} \sim 1,000$~km/s would
almost never occur.  Even for more modest alignments
($\theta_1<30^\circ$), such large kicks only occur for large values of
$\chi$.  Recoils are even smaller for $q < 9/11$, as all components of
the kick are proportional to $q^2/(1+q)^5$ (see Eq.~(11) of
\cite{2012arXiv1201.1923L}).  Note that this kick suppression only
applies when on average $\theta_1 < \theta_2$; for initial values of
$\theta_1 > \theta_2$ the recoil is enhanced, not suppressed, as
demonstrated by the lower two panels of Fig.~\ref{fig:P1000}, where
$\theta_1\geq 150^\circ$. Interestingly, the kick enhancement appears
to saturate: while ejection probabilities approach zero as we fix
$\theta_1$ at ever smaller values, they are essentially independent of
$150^\circ \le \theta_1 \le 175^\circ$ if we apply the kick formula
close to merger. Figure~\ref{fig:P1000} also indicates that kick
suppression via alignment ($\theta_1\sim 0^\circ$) is more efficient
than enhancement via antialignment ($\theta_1\sim 180^\circ$) . We
shall quantify this asymmetry in more detail below.

It may seem surprising that PN spin alignment so dramatically
suppresses the recoil velocities predicted by the new LZ kick formula,
since many of the spin-orbit resonances depicted in Figure~1 of
\cite{Kesden2010} have values of $\theta_i$ comparable to that of the
new ``maximum recoil'' configuration in the presence of the hang-up
effect.  We can understand the effectiveness of PN kick suppression by
recognizing that it results from the alignment of the perpendicular
components of the SBH spins {\it with each other} ($\Delta \phi \to
0^\circ$), not from the spins aligning with the orbital angular
momentum ($\theta_i \to 0$, as Bogdanovi\'{c} et
al. \cite{Bogdanovic2007} argued would occur due to the torque exerted
by a warped circumbinary disk).

To illustrate this point, we consider maximally spinning ($\chi_1 =
\chi_2 = 1$) binaries with $q=9/11$ in Figure~\ref{fig:deltaphi}.
Dotted histograms refer to recoil distributions computed before the PN
evolutions (at $r_i=1000M$), and solid histograms refer to the
corresponding distributions computed after evolving the binaries down
to $r_f=10M$.  Let us first focus on the left panel.  The black curves
in the upper histogram show that black holes with isotropic spin
distributions (flat distributions in $\cos\theta_1$, $\cos\theta_2$,
and $\Delta\phi$) maintain these isotropic distributions as they
inspiral down to $r_f=10M$.  The thick blue (thin red) curves in the
upper histogram corresponds to the subset of this isotropic
distribution with the 30\% lowest (highest) initial values of
$\theta_1$.  Careful comparison of the solid and dotted curves reveals
that recoils are suppressed (enhanced) for low (high) initial values
of $\theta_1$.  This tendency is seen much more clearly in the middle
and lower panels, where $\theta_1$ has the indicated initial value
while $\boldsymbol{\chi}_2$ retains its isotropic initial
distribution.  If $\mathbf{S}_1$ is partially aligned with
$\mathbf{L}$, as in the middle histograms, then $\theta_{12}$ (the
angle between the two spins) and $\Delta\phi$ will be strongly peaked
around $0^\circ$ at the end of the evolution and kicks will be
reduced.  If instead $\mathbf{S}_1$ is partially antialigned with
$\mathbf{L}$ (as in the lower histograms), then $\theta_{12}$ and
$\Delta\phi$ will be strongly peaked around $180^\circ$ at the end of
the evolution and the kicks will be enhanced (cf. the discussion in
\cite{Kesden2010a}).  The crucial element for producing alignment here
is the fact that $\Delta\phi\to 0^\circ$.  This is illustrated in the
right-hand panels of Figure~\ref{fig:deltaphi}, which are the same as
the corresponding left-hand panels, except that $\Delta \phi$ has been
reset to a flat distribution $\Delta \phi \in [0, 180^\circ]$ after
the PN inspiral (but before the recoil velocities are computed).  This
nearly eliminates the kick suppression (for initially aligned
binaries) or enhancement (for initially anti-aligned binaries).

\section{Astrophysical Parameters} \label{S:param}

We demonstrated in the previous section that the new fitting formulae
for gravitational recoils provided in the LZ papers
\cite{Lousto2011,2012arXiv1201.1923L} did not alter our earlier
conclusion \cite{Kesden2010a} that PN spin alignment can dramatically
affect the distribution of recoil velocities for specific initial
values of $q$ and $\theta_1$ and an initially isotropic distribution
for $\boldsymbol{\chi}_2$.  In this section, we describe a more
astrophysically motivated choice for these initial parameters.  The
effect of PN spin alignment on this new distribution will be presented
in Section \ref{S:res}.

\subsection{Mass Ratio} \label{SS:massrat}

The origin of SBHs is poorly understood theoretically and poorly
constrained observationally.  Lynden-Bell \cite{1969Natur.223..690L}
recognized that the central objects of mass $10^7 - 10^9 M_\odot$
believed to power quasars would quickly collapse into SBHs.  Haehnelt
and Rees \cite{1993MNRAS.263..168H} explained the quasar luminosity
function by assuming that such massive SBH seeds promptly formed at
the centers of dark-matter (DM) halos whose formation rates could be
predicted using the Press-Schechter formalism
\cite{1974ApJ...187..425P}.  More recent theoretical work has called
into question whether the high-redshift seeds of SBHs were truly this
massive.  While some maintain that massive central concentrations of
gas at the centers of pre-galactic disks can directly collapse into
$\sim 10^5 M_\odot$ SBH seeds \cite{2006MNRAS.371.1813L}, others argue
that metal-free gas will fragment into Population III stars with
characteristic masses $\gtrsim 100 M_\odot$
\cite{1999ApJ...527L...5B}.  Volonteri, Haardt, and Madau
\cite{Volonteri2003}, also using the Press-Schechter formalism to
estimate halo mass functions and merger rates, showed that
$150~M_\odot$ seeds occupying $3.5-4 \sigma$ overdensities at redshift
$z = 20$ could grow into SBHs that would reproduce both the quasar
luminosity function at $1 \leq z \leq 5$ and the observed local
$M_{\rm BH}-\sigma$ relation \cite{Gebhardt2000} between SBH mass and
galaxy velocity dispersion.  A future space-based interferometer such
as LISA or the European New Gravitational Wave Observatory (NGO/eLISA)
will be able to distinguish between these two different scenarios for
SBH formation by observing the gravitational waves emitted during
high-redshift mergers
\cite{2011MNRAS.415..333P,Gair:2010bx,Sesana:2010wy,2012arXiv1201.3621A}.

Given this uncertainty in the high-redshift population of SBH seeds,
in this paper we restrict our attention to SBH mergers in the
low-redshift ($z \lesssim 1$) universe.  Stewart et
al. \cite{Stewart:2008ep} used a high-resolution N-body simulation to
calculate the merger rates of DM halos in the redshift range $0 \leq z
\leq 4$.  They fit their results for the rate at which a larger galaxy
of mass $M_g$ merges with smaller galaxies with masses between $m_g$
and $M_g$ to the function
\begin{equation} \label{E:mrate}
\frac{dN_{\rm merg}}{dt}(M_g, m_g) = A_t(M_g) F(m_g/M_g)\,,
\end{equation}
where the normalization $A_t(M_g)$ is binned by mass, and the
mass-ratio dependence is given by
\begin{equation} \label{E:mrat}
F(m_g/M_g) = \left( \frac{m_g}{M_g} \right)^{-c} \left( 1 -
\frac{m_g}{M_g} \right)^d\,,
\end{equation}
with the indices $c$ and $d$ determined from the simulation.  This
same functional form can be used for the merger rates of both DM halos
of mass $M_h$ and galaxies with stellar mass $M_\ast$, although the
fitted values of the parameters will be different.  Table 1 of Stewart
et al.  \cite{Stewart:2008ep} provides numerical estimates of these
parameters binned by stellar mass; we use the functions
\begin{subequations} \label{E:mparam}
\begin{eqnarray}
A_t(M_\ast) &=& 0.0098~{\rm Gyr}^{-1} 
\left( \frac{M_\ast}{10^{10}~M_\odot} \right)^{0.736}\,, \\
\label{E:Anorm}
c &=& 0.329 \left( \frac{M_\ast}{10^{10}~M_\odot} \right)^{-0.158}\,, \\
\label{E:cgal}
d &=& 1.1 - 0.2z\,,
\label{E:dgal}
\end{eqnarray}
\end{subequations}
which approximate the values listed in this table.  Note that the
total rate of mergers diverges in the limit $m_g \to 0$ for $c > 0$,
although the rate at which the galactic mass increases remains finite.
The Press-Schechter formalism also predicts that the number of DM
halos and thus the number of mergers diverges as their mass goes to
zero.  Volonteri, Haardt, and Madau \cite{Volonteri2003} address this
issue in their merger trees by only keeping track of halos with masses
above an effective mass resolution $M_{\rm res}$ that evolves with
redshift, but always remains a small fraction of the largest
progenitor mass.  Because we are only interested in SBH mergers at low
redshift, we neglect galaxies with $M_\ast < 10^9 M_\odot$.  This
corresponds to neglecting SBHs with $m_i < 10^6 M_\odot$, since SBHs
are about a fraction of $10^{-3}$ of the stellar mass of their host
spheroids \cite{Merritt2001}.  SBHs less massive than this value are
poorly constrained observationally, and even if present in a halo may
have difficulty reaching the galactic center, since the dynamical
friction time scales as the inverse of the SBH mass.  We assume that
SBH mergers occur promptly following the merger of their host
galaxies.

Eq.~(\ref{E:mrate}) gives the rate at which individual galaxies of
stellar mass $M_\ast$ merge with smaller galaxies; to determine the
total rate of mergers in a cosmological volume also requires an
estimate of the galactic stellar mass function $dN_g/d\log M_\ast$.
We use the mass function determined from the $z \sim 0.37$ sample of
the COSMOS survey \cite{Leauthaud:2011rj}; this mass function, as
shown in the left panel of Figure~14 of \cite{Leauthaud:2011rj}, is
quite similar to earlier mass functions derived from the Sloan Digital
Sky Survey
\cite{2007MNRAS.378.1550P,2008MNRAS.388..945B,2009MNRAS.398.2177L}.
The total rate per unit cosmological volume $R(M_\ast)$ at which
galaxies with stellar masses between $10^9~ M_\odot$ and $M_\ast$
merge is given by
\begin{equation} \label{E:mergrate}
R(M_\ast) = \int_{9}^{\log M_\ast} \frac{dN_g}{d\log M_\ast} \frac{dN_{\rm merg}}{dt}(M_{\ast}^{\prime},
10^9 M_\odot)~d\log M_{\ast}^{\prime}~.
\end{equation}

We use this rate function as part of a Monte Carlo approach to
generate a representative sample of SBH mergers.  We choose a random
number $r_1$ from a flat distribution in the interval $0 \leq r_1 \leq
1$, then find the stellar mass $M_\ast$ for which $R(M_\ast) = r_1
R(\infty)$.  Values of $r_1 \simeq 0$ correspond to stellar masses
$M_\ast \simeq 10^9 M_\odot$, while values $r_1 \simeq 1$ correspond
to stellar masses $M_\ast \simeq 10^{12} M_\odot$ above which the
galactic luminosity function is exponentially suppressed.  We next
choose a second random number $r_2$, also from a flat distribution $0
\leq r_2 \leq 1$, and find the value of $m_\ast$ for which
$F(m_\ast/M_\ast) = r_2 F(10^9 M_\odot/M_\ast)$.  Values of $r_2
\simeq 0$ correspond to nearly equal-mass mergers ($m_\ast \lesssim
M_\ast$), while values $r_2 \simeq 1$ correspond to mergers with the
smallest galaxies to host SBHs ($m_\ast \simeq 10^9 M_\odot$).  The
SBH masses are then determined from the masses of their host galaxies,
$m_1 = 10^{-3} M_\ast$ and $m_2 = 10^{-3} m_\ast$.  Histograms of our
SBH mass ratio distribution are given by the solid black curves in the upper
panels of Fig.~\ref{fig:eject1000log}.

Our assumption that the SBH mass is $10^{-3}$ that of its host galaxy's stellar
mass is strictly true only for elliptical galaxies, since observations suggest
that SBH masses are uncorrelated with galactic disks and pseudobulges
 \cite{2011Natur.469..374K}.  The most massive galaxies are primarily ellipticals,
 while smaller galaxies are a mixture of ellipticals and spirals \cite{2003ApJ...594..186B}.
 A more sophisticated Monte Carlo treatment would draw galactic bulge fraction
 from an observed distribution as a function of galactic mass, and only correlate
 the SBH mass with this bulge component.  Such a treatment is beyond the scope
 of this paper.

This procedure for determining the SBH mass ratio $q = m_2/m_1$ is
considerably more elaborate than that used in the LZ papers
\cite{Lousto2011,2012arXiv1201.1923L}.  They assumed that the
probability of having a SBH mass ratio between $q$ and $q + dq$ was
simply $F(q) dq$, with the function $F(q)$ given by Eq.~(\ref{E:mrat})
with $c = 0.3$ and $d = 1$.  This assumption seems to be based on a
misinterpretation of Stewart et al. \cite{Stewart:2008ep}, which
defines $F(q)$ as being proportional to the total number of mergers
between a mass ratio of $q$ and 1.  The LZ papers did not need to
introduce an explicit cutoff at small mass ratios, because the total
number of mergers remained finite under their assumption.  This finite
number of mergers allowed them to calculate the probability that the
gravitational recoil would be in a particular velocity interval.
Given the infinite rate of mergers predicted by Eq.~(\ref{E:mrate}) as
$m_\ast \to 0$, we would predict an infinite number of mergers with $v
= 0$~km/s in the absence of a cutoff, since $v \propto q^2$ as $q \to
0$.  This cutoff is well justified physically because DM halos of
arbitrarily small mass cannot host stars, let alone SBHs.

\subsection{Initial Spins} \label{SS:spins}

The accretion and merger history of SBHs determines the magnitude and
direction of their spins. In this section we discuss some
astrophysical predictions on SBH spin magnitude and direction.

\subsubsection{Spin Magnitudes} \label{SSS:mag}

Here we present a short overview of the literature to show that SBH
spin magnitudes are strongly dependent on the underlying assumptions
about gas accretion. Because of these uncertainties, in this paper we
simply assume $\chi_1 = \chi_2 = \chi$ for $\chi = 0.5$, 0.75, and
1.0, and we provide predictions for each of these three values.
Schnittman \cite{Schnittman2004} showed that the amount of PN spin
alignment is insensitive to the spin magnitude for $\chi \gtrsim 0.5$
(see Fig. 11 in his paper).  At large separations where $|\mathbf{L}|
\gg |\boldsymbol{\chi}_i|$, Eq.~(3.5) of \cite{Schnittman2004}
indicates that the constraint for defining a spin-orbit resonance
depends only on $\theta_i$, not $\chi_i$.  We therefore expect that PN
spin alignment should not be particularly sensitive to our assumption
that $\chi_1 = \chi_2$, although additional numerical studies would be
required to confirm this expectation.

Volonteri et al. \cite{Volonteri2004} examined the distribution of SBH
spin magnitudes using cosmological merger trees constructed within the
Press-Schechter formalism.  They found that binary mergers alone lead
to broad distributions of the dimensionless spin magnitude peaked
around $\chi \simeq 0.5$, but that when accretion is included the
spins reach much larger values.  Accretion from geometrically thin
disks \cite{1973A&A....24..337S} leads to spin distributions sharply
peaked around $\chi \simeq 0.998$ \cite{1974ApJ...191..507T}, as the
Bardeen-Petterson effect \cite{1975ApJ...195L..65B} aligns the SBH
spins with the disks on a timescale \cite{2009MNRAS.399.2249P}
 \begin{equation} \label{E:tal}
 t_{\rm al} \sim 10^5 \chi^{5/7} \left( \frac{M}{10^6 M_\odot} \right)^{-2/35} \left( \frac{f_{\rm Edd}}{\eta_{0.1}}
 \right)^{-32/35}~{\rm yr}\,,
 \end{equation}
much shorter than the Salpeter time
 \begin{equation} \label{E:tSal}
 t_{\rm Edd} \simeq 4.6 \times 10^7 \left( \frac{\eta_{0.1}}{f_{\rm Edd}} \right)~{\rm yr}
\end{equation}
on which the spin magnitudes change.  Here $\eta_{0.1}$ is the
radiative efficiency normalized by a typical value $0.1$ and $f_{\rm
  Edd}$ is the AGN luminosity in units of the Eddington value $L_{\rm
  Edd} = 4\pi G M_{\rm BH} c m_p/\sigma_T$ ($m_p$ is the proton mass,
$\sigma_T$ is the Thomson cross section for elastic scattering).
Accretion from geometrically thick disks leads to much broader spin
distributions, because spin alignment occurs on the longer Salpeter
time.  SBHs that accrete from a thick disk may also have a smaller
maximum spin $\chi \simeq 0.93$, as ordered magnetic fields in the
plunging region interior to the innermost stable orbit may extract
angular momentum that would be advected by the SBH in the thin-disk
case \cite{2004ApJ...602..312G}.

Volonteri et al. \cite{Volonteri2004} assumed that after each major
merger, SBHs accreted continuously from disks with well defined
angular momentum until their mass increased by an amount $\Delta m =
3.6 \times 10^6 V_{c, 150}^{5.2} M_\odot$, where $V_{c, 150}$ is the
circular velocity of the host galaxy's DM halo in units of 150~km/s.
Moderski and Sikora \cite{1996A&AS..120C.591M} proposed an alternative
scenario, in which gas was accreted in small discrete episodes with
random orientations with respect to the SBH spin.  This ``chaotic
accretion'' scenario leads to SBH spins that fluctuate about $\chi =
0$, and was originally motivated by the conjecture of Wilson and
Colbert \cite{1995ApJ...438...62W} that only radio-loud AGN
(approximately 10\% of all AGN) are powered by highly spinning SBHs.  King et al.
\cite{King2008} suggested that the mass accreted in each of these
discrete episodes might be determined by the mass of the accretion
disk interior to the radius $r_{\rm sg} \sim 0.01 - 0.1$~pc at which
the disk becomes self-gravitating.  Star formation at $r > r_{\rm sg}$
would stir the gas flow, implying that each event would have
essentially random orientations.  Recent high-resolution N-body/SPH
simulations of galactic accretion disks which include star formation
support this suggestion \cite{2011arXiv1111.1236H}.  These simulations
follow gas flows from galactic scales of $\sim 100$~kpc all the way
down to $< 0.1$~pc, and show that the orientation of the inner nuclear
disk varies on $\sim$~Myr timescales, that are comparable to the
alignment timescale $t_{\rm al}$ of Eq.~(\ref{E:tal}).

Berti and Volonteri \cite{Berti2008} explored how chaotic accretion
would affect SBH spin evolution in a cosmological context, also
incorporating the results of NR simulations available at the time.
The NR simulations implied that SBH spin distributions would now be
broadly peaked about $\chi \sim 0.7$ in the absence of accretion, but
that the full range of spin magnitudes $0 \leq \chi \leq 1$ was now
possible once both standard and chaotic accretion were permitted.
This differs from the spin magnitude distributions used in the LZ
papers \cite{Lousto2011,2012arXiv1201.1923L}, which as shown in
Figure~7 of \cite{2012arXiv1201.1923L} are peaked around $\chi \simeq
0.8$.  These spin distributions were determined from the N-body/SPH
simulations of Dotti et al. \cite{Dotti:2009vz}, which were restricted
to equal-mass binaries, and began with an initially uniform
distribution of spin magnitudes $\chi_i\in [0,1]$.  Although the
$\lesssim 10$~Myr duration of these simulations is long compared to
the alignment time $t_{\rm al}$ on which the spin directions change,
it is less than the Salpeter time $t_{\rm Edd}$ on which the spin
magnitudes change.  One might therefore be concerned that these
distributions have not converged to the value they should have at
$r_i\sim 1000M$, where the gravitational-wave driven stage of the
inspiral begins.

\subsubsection{Spin Directions} \label{SSS:dir}

The residual misalignment of the SBH spins with their orbital angular
momentum at $r_i$ has been less thoroughly investigated than the spin
magnitudes. We review some of these studies below. To encompass all
possible scenarios with regard to astrophysical spin alignment prior
to $r_i = 1000M$, in our simulations we will consider:
\begin{itemize}

\item[1)] Uniform symmetric distributions in $\cos \theta_1$ and
  $\cos\theta_2$ drawn in the range $\theta_i\in[0^\circ,10^\circ]$
  ($i=1\,,2$; strong alignment) and $\theta_i\in[0^\circ,30^\circ]$
  (weak alignment). We will refer to these cases as the ``10/10'' and
  ``30/30'' scenarios.
\item[2)] A uniform distribution in $\cos \theta_1$ and $\cos\theta_2$
  with asymmetric range for the angles:
  $\theta_1\in[0^\circ,10^\circ]$, $\theta_2\in[0^\circ,30^\circ]$. In
  this ``10/30'' scenario the primary is more strongly aligned with
  the orbital angular momentum.
\item[3)] A uniform distribution in $\cos \theta_1$ and $\cos\theta_2$
  with $\theta_1\in[0^\circ,30^\circ]$,
  $\theta_2\in[0^\circ,10^\circ]$. This ``30/10'' scenario could
  result from the secondary orbiting close to the inner edge of a
  circumbinary disk, so that it ends up being more aligned than the
  primary with the orbital angular momentum.
\end{itemize}

In order to justify these assumptions, here we review some literature
on astrophysical predictions for the spin orientation resulting from
SBH mergers.  

Dotti et al. \cite{2009MNRAS.396.1640D} performed simulations where
one $4 \times 10^6 M_\odot$ SBH is initially at rest at the center of
a $10^8 M_\odot$ circumnuclear disk.  The second SBH, also with $m_i =
4 \times 10^6 M_\odot$, began at an initial separation of $r = 50$~pc
with orbital angular momentum that was already aligned or anti-aligned
with that of the circumnuclear disk.  This is consistent with the SPH
simulations of Larwood and Papaloizou \cite{1997MNRAS.285..288L},
which suggest that the mass-quadrupole moment of the SBH binary will
induce differential precession in an inclined circumbinary disk.
Viscous dissipation will then cause this differentially precessing
disk to settle into the equatorial plane of the binary.  It is unclear
whether this mechanism will operate at the large initial separation of
the Dotti et al. simulations before the SBH binary becomes
gravitationally bound.  Dynamical friction causes initially eccentric
or retrograde orbits to circularize as the second SBH inspirals to a
final separation $r \simeq 5$~pc.  Accretion during this inspiral
aligns the spins of both SBHs to within $10^\circ$ of the orbital
angular momentum for a cold circumnuclear disk, and to within
$30^\circ$ for a hot disk with greater pressure support, and hence more
isotropic gas velocity dispersion \cite{Dotti:2009vz}.  These
simulations did not include star formation or cooling, and the
internal energy of the gas particles was chosen to prevent
gravitational fragmentation.  These choices suppress chaotic accretion
and the dynamical instabilities found in Hopkins et
al. \cite{2011arXiv1111.1236H}, that cause the direction of the
angular momentum of the inner disk to fluctuate on $\lesssim$~Myr
timescales.  These estimates should therefore be considered lower
bounds on the residual misalignment between the SBH spins and their
orbital angular momentum.

After the simulations of Dotti et al. \cite{2009MNRAS.396.1640D} end,
but before the gravitational-wave driven stage of the inspiral begins
at $r_i = 1000 M$, the SBHs open a gap and form a true circumbinary
disk.  High-resolution hydrodynamical simulations indicate that the
gas surface density will be sharply truncated at radii less than twice
the semi-major axis of the binary \cite{2008ApJ...672...83M}.  Mass
can flow from the inner edge of this circumbinary disk onto the
individual SBHs \cite{1996ApJ...467L..77A,2007PASJ...59..427H}.
Accretion rates are generally higher for the less massive SBH, which
is further from the center of mass and hence closer to the inner edge
of the disk.  This point is absolutely crucial for the PN spin
alignment that is the subject of this paper, as it is the {\it
  relative} value of $\theta_1$ and $\theta_2$ that determines whether
gravitational recoils are suppressed or enhanced.  One might expect
that the higher accretion rate onto the smaller SBH would lead to
closer alignment between its spin and that of the orbital angular
momentum.  Alternatively, it is possible that the spin of the more
massive SBH would be more influential in determining the direction of
the warp in the circumbinary disk that torques the SBH spins.  The
relative value of $\theta_1$ and $\theta_2$ in the presence of a
circumbinary disk remains very much an open question.

\begin{figure*}[thb]
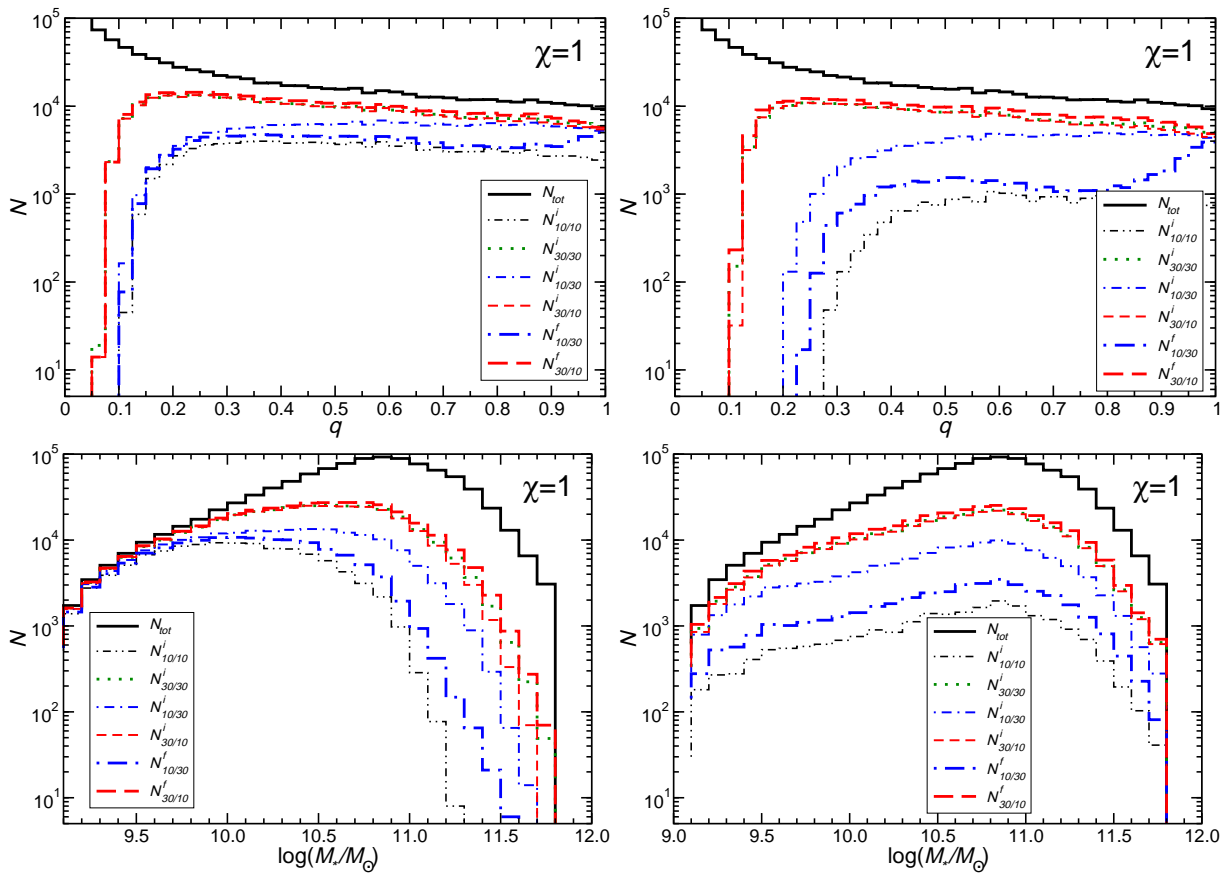

\begin{center}
\epsfig{clip=true,file=fig3a.eps,width=8.0cm}
\epsfig{clip=true,file=fig3b.eps,width=8.0cm}
\epsfig{clip=true,file=fig3c.eps,width=8.0cm}
\epsfig{clip=true,file=fig3d.eps,width=8.0cm}
\end{center}
\caption{Distributions of ejected remnants as functions of the binary
  mass ratio (top) and of the mass of the host galaxy (bottom).  Left
  panels: $v_{\rm ej}$ is computed using
  Eq.~(\ref{vescmerritt}). Right: we use a constant ejection velocity
  $v_{\rm ej}=1,000$~km/s. In the asymmetric cases (10/30 and 30/10)
  thin lines refer to kicks computed using the initial spin
  distributions, while thick lines refer to kicks computed from the
  final distributions.}
\label{fig:eject1000log}
\end{figure*}

A final complication is the growth of the binary's orbital
eccentricity through its interaction with the circumbinary disk
\cite{2001A&A...366..263P}.  Although dynamical friction damped the
orbital eccentricity at the large separations of the Dotti et
al. simulations \cite{2009MNRAS.396.1640D}, after the formation of a
circumbinary disk the orbital eccentricity again grows to a limiting
value $e_{\rm crit} \simeq 0.7$ \cite{2011MNRAS.415.3033R}.
Gravitational radiation circularizes the binary after it decouples
from the circumbinary disk \cite{Peters1964}, but the binary could
still have considerable residual eccentricity $e_{\rm LISA} \propto
M^{-0.73} q^{-1.2}$ when the frequency of the fundamental GW harmonic
reaches $f_{\rm LISA} = 10^{-4}$~Hz
\cite{2011MNRAS.415.3033R,Yunes:2009yz}.
We will neglect residual eccentricity during the inspiral and any
effects it might have on PN spin alignment.  

\section{Results} \label{S:res}

We carried out a systematic study of the probability that
gravitational recoils eject SBHs from their host galaxies.  We
generated a population of binaries with astrophysically motivated mass
ratios using the Monte Carlo prescription described in
Section~\ref{SS:massrat}.  To quantify the effect of SBH spin
magnitudes, we computed ejection probabilities for three fixed values
of $\chi_1=\chi_2=\chi$: $\chi$=0.5, 0.75, and 1.  To account for all
possibilities given the great uncertainty in the spin-alignment
distributions, we considered the four cases described in
Section~\ref{SSS:dir}: two ``symmetric'' cases (10/10, 30/30) and two
``asymmetric'' cases (10/30 and 30/10).

Our results for $\chi=1$ are summarized in
Figure~\ref{fig:eject1000log}.  This figure has four panels: our
simulated binaries are binned by mass ratio $q$ in the top panels, and
by the stellar mass $M_\ast$ of the larger host galaxy in the bottom
panels.  In the right panels we assume that a gravitational recoil
greater than $v_{\rm ej}=1,000$~km/s is needed to eject a recoiling
SBH from its host galaxy, while in the left panels we adopt a
prescription that depends on the stellar mass $M_\ast$.  The
gravitational potential of a real galaxy is the sum of contributions
from its stellar mass and DM halo; although the DM halo is more
massive, the more concentrated stellar contribution is dominant for
large elliptical galaxies \cite{Merritt2004}.  Modeling elliptical
galaxies with the density profiles of Terzi\'{c} and Graham
\cite{Terzic:2005hy}, Gualandris and Merritt \cite{Gualandris2007}
find that escape velocities have typical values $v_{\rm esc} \simeq
2.1 (GM_\ast/R_e)^{1/2}$, where $R_e$ is the effective radius that
contains half the galaxy's total luminosity.  Combining this with the
observed relation $R_e \approx 1.2~{\rm kpc} (M_\ast/10^{10}
M_\odot)^{0.075}$ yields the approximate expression
\cite{Gualandris2007}
\begin{equation} \label{vescmerritt}
v_{\rm esc} \simeq 1,154~{\rm km/s} 
\left( \frac{M_\ast}{10^{11} M_\odot} \right)^{0.4625}~.
\end{equation}
This is the prescription we used to calculate escape velocities in the
left panels of Figure~\ref{fig:eject1000log}.

Each panel of Figure~\ref{fig:eject1000log} has 7 histograms: the
solid black curve corresponds to the total distribution of simulated
binaries, while the 6 dashed and/or dotted colored curves correspond
to the distributions of ejected SBHs given the 6 different
spin-alignment distributions indicated in the keys to the figure.  PN
spin alignment can play an important role for asymmetric
distributions, as can be seen by comparing the distributions of
ejected binaries calculated from spins {\it before} (thin-line dashed
and dash-dotted histograms) or {\it after} (thick-line histograms) the
PN evolution from $r_i=1000M$ to $r_f=10M$. As expected, the fraction
of ejected binaries decreases in the 10/30 case, while it increases in
the 30/10 case.  However, the kick reduction in the 10/30 case is more
significant than the kick enhancement in the 30/10 case: in other
words, {\em PN resonances are more effective at reducing recoils when
  the primary is more aligned with the orbital angular momentum than
  they are at increasing recoils when the secondary is more aligned
  with the orbital angular momentum}.  For the symmetric distributions
(10/10 and 30/30), PN spin alignment has a marginal effect on the
ejection probabilities\footnote{We have checked this statement by
  running PN evolutions in the 10/10 case. The ``initial'' and
  ``final'' distributions of ejected merger remnants are very
  similar.}.  For clarity (and to save computational time), in these
symmetric cases we only plot the distribution of ejected binaries that
results by applying the recoil formula to the initial distribution at
$r_i=1000M$.

Comparing the left and right panels of Figure~\ref{fig:eject1000log}
allows us to understand the effects of the mass-dependent escape
velocity of Eq.~(\ref{vescmerritt}) on the ejection probability.  The
most obvious effect is the steep decrease in the ejected fraction from
near unity at $M_\ast \lesssim 10^{10} M_\odot$ to less than 10\% for
$M_\ast \gtrsim 10^{11} M_\odot$.  This is very natural: heavier
galaxies should be more effective at retaining recoiling SBHs.
Somewhat more surprising is that despite the constant escape velocity,
the escape fraction also increases as $M_\ast \to 10^9 M_\odot$ in the
bottom right panel.  This is a consequence of our decision to neglect
SBHs with $m_2 < 10^6 M_\odot$ that reside in galaxies with $m_\ast <
10^9 M_\odot$.  Because of this choice, SBHs in galaxies with $M_\ast
\simeq 10^9 M_\odot$ only undergo comparable-mass ($q \simeq 1$)
mergers, elevating the fraction of large recoils since kicks are
proportional to $q^2/(1+q)^5$ at leading order (see Eq.~(11) of
\cite{2012arXiv1201.1923L}).

\begin{figure*}[htb]
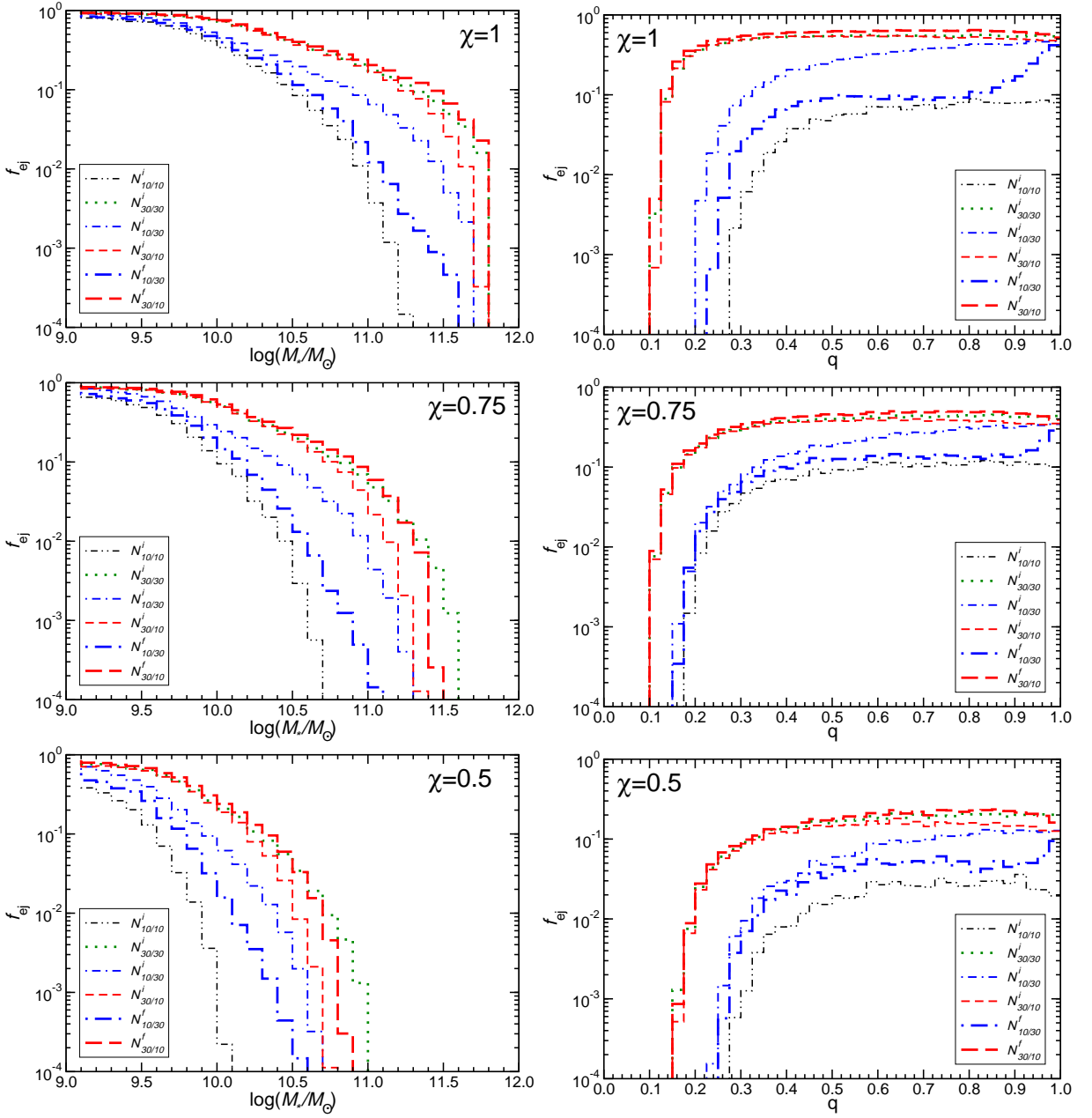

\begin{center}
\epsfig{clip=true,file=fig4a.eps,width=8.0cm}
\epsfig{clip=true,file=fig4b.eps,width=8.0cm}
\epsfig{clip=true,file=fig4c.eps,width=8.0cm}
\epsfig{clip=true,file=fig4d.eps,width=8.0cm}
\epsfig{clip=true,file=fig4e.eps,width=8.0cm}
\epsfig{clip=true,file=fig4f.eps,width=8.0cm}
\end{center}
\caption{Fraction $f_{\rm ej}$ of ejected remnants as functions of the mass of the
  host galaxy (left) and of the binary mass ratio (right).  Here
  $v_{\rm ej}$ is computed using Eq.~(\ref{vescmerritt}) and we vary
  the dimensionless spin parameter $\chi=\chi_1=\chi_2$.}
\label{fig:ejectratio}
\end{figure*}

A second effect of the mass-dependent escape velocity of
Eq.~(\ref{vescmerritt}) is the possibility of ejecting recoiling SBHs
for mass ratios as low as $q \simeq 0.1$, as can be seen in the top
left panel of Figure~\ref{fig:eject1000log}.  The larger host galaxy
can have a mass as small as $10^{10} M_\odot$ for these mass ratios,
corresponding to an escape velocity $v_{\rm esc} \simeq 400~{\rm km/s}
< 1,000$~km/s by Eq.~(\ref{vescmerritt}).  This accounts for the
nonzero escape probabilities for $q < 0.1$.  We also note that in
preparing these figures we have not evolved any PN inspirals for $q <
0.1$, even for those histograms labeled with $N^f$ that correspond to
evolved distributions.  This explains why the thin and thick lines
corresponding to the initial and final histograms coincide for $q <
0.1$.  The gravitational-wave inspiral time is inversely proportional
to the binary's symmetric mass ratio $\eta=q/(1+q)^2$
\cite{Peters1964}, making it hard to preserve accuracy over long PN
time evolutions for $q < 0.1$.  However, Schnittman
\cite{Schnittman2004} showed that spin-orbit resonances only become
important inside a ``resonance locking'' radius $r_{\rm lock}/M
\approx [(1+q^2)/(1-q^2)]^2$.  We are therefore justified in treating
all binaries with $q\leq 0.1$ (for which PN evolutions from
$r_i=1000M$ to $r_f=10M$ become problematic) as if the inspiral did
not happen: that is, we can compute recoils of binaries with $q<0.1$
by simply applying the recoil formula at $r=r_i$.

We examine how the fraction $f_{\rm ej}$ of ejected SBHs depends on
the SBH spin magnitude $\chi$ in Figure~\ref{fig:ejectratio}.  The two
top panels in this figure were produced using the same distributions
shown in the left panels of Figure~\ref{fig:eject1000log}; the
fraction $f_{\rm ej}$ is calculated by dividing the histograms for
each spin distribution by the solid black histogram showing the total
distribution.  As expected, the ejected fractions are smaller for
lower values of $\chi$, since the kick magnitudes are reduced while
the escape velocities remain fixed.  Ejections from host galaxies with
$M_\ast \gtrsim 10^{11.5} M_\odot$ ($10^{11} M_\odot$) are nearly
eliminated for $\chi \lesssim 0.75$ (0.5).  PN spin alignment remains
effective for these lower spin magnitudes; for certain values of
$M_\ast$, the ejected fraction is reduced by almost an order of
magnitude.  Careful scrutiny of the left panels of
Figure~\ref{fig:ejectratio} reveals that PN kick enhancement for the
30/10 case drives $f_{\rm ej}$ above that of the 30/30 case, except at
the highest values of $M_\ast$ for the $\chi = 0.5$ and 0.75
distributions.  Mergers involving the most massive host galaxies are
dominated by small values of $q$, for which PN spin alignment is
ineffective.

It is also interesting to note that PN spin alignment is suppressed as
$q \to 1$ (the thin and thick lines converge in this limit in the
right panels of Figure~\ref{fig:ejectratio}).  Schnittman
\cite{Schnittman2004} showed that PN spin alignment increased as $q$
increased from 1/9 up to 9/11 (see Figure~10 of
\cite{Schnittman2004}).  Symmetry implies that PN spin alignment must
vanish for equal-mass binaries, since the labeling of the SBHs is
arbitrary for $q = 1$ and it is therefore impossible to distinguish
$\theta_1 < \theta_2$ from $\theta_1 > \theta_2$.  This limiting
behavior implies that the effects of spin alignment must be maximized
for some mass ratio in the range $0.8 \leq q \leq 1$, which is
precisely what we observe in the right panels of
Figure~\ref{fig:ejectratio}.  These near-unity mass ratios are also
expected to yield the largest kicks, which emphasizes the importance
of including the effects of PN spin alignment in future studies of
gravitational recoils.

A final summary of our results is provided in Table~\ref{T:fej}, where
we calculate the percentage of ejected binaries for each spin
distribution.  Although the precise amount of spin alignment depends
on the value of $\chi$, we see that kicks are suppressed (enhanced) by
$\sim 40\%$ (20\%) when the escape velocities are given by
Eq.~(\ref{vescmerritt}).  The kick suppression or enhancement is even
more dramatic for the case of a constant escape velocity $v_{\rm ej} =
1,000$~km/s.  The ejected fraction is lower overall for this choice of
escape velocity, which is larger than the mean value predicted by
Eq.~(\ref{vescmerritt}).  The kick suppression or enhancement is
greater because the effects of PN spin alignment are most pronounced
at the high--velocity tail of the kick distribution, as shown in the
left panel of Figure~\ref{fig:deltaphi}.

\begin{center}
\begin{table}[t!]
\begin{tabular}{ccccccc}
\hline
\hline
$\chi$ &$(10/10)_i$ &$(30/30)_i$ &$(10/30)_i$ &$(10/30)_f$ &$(30/10)_i$ &$(30/10)_f$\\ 
\hline
\hline
\multicolumn{7}{c}{$v_{\rm ej}$ given by Eq.~(\ref{vescmerritt})}\\
0.50   &0.76    &7.21    &2.95    &1.59    &6.03    &7.81   \\
0.75   &3.86    &19.28   &9.208   &5.310   &17.79   &21.41  \\
1.00   &11.22   &34.26   &19.49   &13.70   &33.60   &37.11  \\
\hline
\multicolumn{7}{c}{$v_{\rm ej}=1,000$~km/s}\\
0.50   &0       &0.39    &0       &0       &0       &0.002  \\
0.75   &0       &10.30   &2.21    &0.13    &7.70    &12.92  \\
1.00   &2.21    &27.57   &12.18   &4.19    &26.87   &31.22  \\
\hline
\hline
\end{tabular}
\caption{\label{tab:bounds} Percentage of ejected binaries for
  different values of the spin and different combinations of the
  angles ($\theta_1$, $\theta_2$). A column header such as
  ``$(10/30)_i$'' means that the probability was computed considering
  the spin distribution before PN evolutions, while ``$(10/30)_f$''
  means that the probability was computed applying the recoil formula
  at the end of the PN evolution.  In the top rows we assume that the
  escape velocity $v_{\rm ej}$ is given by Eq.~(\ref{vescmerritt}); in
  the bottom rows we assume a constant $v_{\rm ej}=1,000$~km/s.}
\label{T:fej}  
\end{table}
\end{center}

\section{Discussion} \label{S:disc}

We demonstrated in previous work \cite{Kesden2010a} that the alignment
of SBH spins with each other during the PN stage of the inspiral can
dramatically suppress the predicted distribution of gravitational
recoils.  This mechanism can reinforce the kick suppression that
results from the alignment of the SBH spins with their orbital angular
momentum through interaction with a circumbinary disk at an earlier
stage of the inspiral \cite{Bogdanovic2007}.  Recent NR simulations
\cite{Lousto2011,2012arXiv1201.1923L} showed that gravitational
recoils are maximized not in the previously claimed ``superkick''
configuration in which the spins lie in the orbital plane, but in a
new ``hang-up'' configuration in which the angle between the spins and
orbital angular momentum is $\sim 50^\circ$.  The primary conclusion
of this paper is that kick suppression due to PN spin alignment
remains highly effective despite the new dependence of the
gravitational recoil on SBH spins.

It is difficult to make very robust quantitative predictions about the
magnitude of this effect, as the recoil distribution is highly
sensitive to SBH spin distributions that are theoretically uncertain
and poorly constrained observationally.  Some SPH simulations have
shown that SBH spins can grow to large magnitudes and rapidly align
with the orbital angular momentum through coherent accretion from a
massive circumnuclear disk \cite{Dotti:2009vz}, but other simulations
that include gas cooling and star formation indicate that the
direction of the angular momentum of the accreted gas will vary
chaotically on timescales comparable to the alignment time, leading to
smaller spin magnitudes and larger misalignments with the orbital
angular momentum \cite{2011arXiv1111.1236H}.  Our results show that
the predicted fraction of recoiling SBHs that are ejected from their
host galaxies can vary from $\lesssim 10^{-2}$ to $\sim 1/3$ depending
on the adopted distribution of SBH spins.  Although somewhat
frustrating from a theoretical perspective, the strong dependence of
the ejected fraction on SBH spins implies that observational studies
of recoils may place promising constraints on the highly elusive SBH
spin distribution.

Our work in this paper reveals several of the issues that must be
addressed before we can predict recoil distributions with confidence:

\begin{itemize}

\item[1)] We need to determine how effectively realistic circumbinary
  disks can align SBH spins with their orbital angular momentum.  Such
  circumbinary disks may not remain geometrically thin, and may
  gravitationally fragment or collapse into stars.  These
  possibilities could have significant quantitative effects on the SBH
  spin distribution at the onset of the PN stage of the inspiral.

\item[2)] We need to develop a better understanding of the interaction
  between unequal-mass SBHs and a circumbinary disk.  One could argue
  that the spin of either the primary or the secondary might align
  more efficiently with the orbital angular momentum.  Whether PN spin
  alignment leads to kick suppression or enhancement depends crucially
  on the {\em relative} values of $\theta_1$ and $\theta_2$.

\item[3)] We need to carefully consider how the merger rate,
  mass-ratio distribution, and escape velocity depend on host-galaxy
  mass.  Our results shown in Figs.~\ref{fig:eject1000log} and
  \ref{fig:ejectratio} demonstrate the sensitivity of the ejected
  fraction to assumptions about these galactic properties.

\end{itemize}

Although challenging, steady theoretical and observational progress is
being made on all of these issues.  Once astrophysicists can provide
more accurate predictions of SBH spin distributions at $r_i \sim
1000M$, when SBHs decouple from their circumbinary disks, relativists
will be able to evolve the binaries through merger and more accurately
predict gravitational recoils.  Including the PN stage of the inspiral
will be an important ingredient in this grand theoretical undertaking.

\vspace{0.3cm}

{\bf \em Acknowledgements.}
We thank Mike Blanton, Massimo Dotti, Mike Eracleous, David Hogg, Carlos Lousto, David Merritt,
Jeremy Tinker, Marta Volonteri, and Yosef Zlochower for useful
conversations.  This work was supported by NSF Grants PHY-090003,
PHY-0900735 and PHY-1055103, FCT projects PTDC/FIS/098032/2008 and
PTDC/FIS/098025/2008, the Ram{\'o}n y Cajal Programme and Grant
FIS2011-30145-C03-03 of the Ministry of Education and Science of
Spain, the FP7-PEOPLE-2011-CIG Grant CBHEO, No.~293412, the Sherman
Fairchild Foundation to Caltech, projects ICTS-CESGA-221 of the Centro
de Supercomputaci{\'o}n de Galicia, AECT-2011-3-0007 by the Barcelona
Supercomputing Center, and the ERC Starting Grant ERC-2010-Stg-DyBHo.

%

\end{document}